\documentstyle[12pt]{article}

\begin{document}
\bibliographystyle{unsrt}
\def\tm {\tilde m^2}
\def\tl {\tilde\lambda}

\begin{flushright}
SINP-TNP/94-12\\
TIFR/TH/94-37\\
\end{flushright}

\begin{center}
{\Large {\bf TAMING THE SCALAR MASS PROBLEM WITH A SINGLET
HIGGS BOSON}}\\[5mm]
Anirban Kundu\\ {\em Theory Group, Saha Institute of Nuclear Physics,\\
1/AF Bidhannagar, Calcutta - 700064, India.\\
e-mail: akundu@saha.ernet.in}\\[1mm]
and\\[1mm]
Sreerup Raychaudhuri\\ {\em Theoretical Physics Group,\\
Tata Institute of Fundamental Research,\\ Homi Bhabha Road, Bombay -
400005, India.\\
e-mail: sreerup@theory.tifr.res.in}\\[10mm]
\end{center}
\begin{abstract}

We investigate the fine-tuning problem in the Standard Model
and show that Higgs boson and top quark masses consistent with
current experimental bounds cannot be obtained unless one
extends the particle spectrum. A minimal extension
which achieves this involves addition of
a singlet real scalar and one generation of vectorlike fermions.
We show that this leads to a phenomenologically viable
prediction for the mass of the Standard Model Higgs boson.

\end{abstract}
\vskip 20pt

October 1994

\newpage

\begin{center}
{\Large{\bf 1. Introduction}}
\end{center}
\bigskip

Electroweak precision tests at the CERN $e^+ e^-$ collider LEP
and the recently reported discovery of the top
quark at Fermilab \cite{abe} have established beyond reasonable doubt
the fact that the Standard Model (SM) is an excellent description
of fundamental interactions at least upto the electroweak
symmetry-breaking scale. Nevertheless, there is a general belief
that the SM does not tell us the whole story, but merely provides
an effective Lagrangian of a deeper underlying theory which is yet
to be established. One of the chief reasons for such a belief is
the so-called {\it fine-tuning} problem.

In a nutshell, the fine-tuning problem is the following. The masses
of scalars --- specifically the Higgs boson --- receives radiative
corrections which are quadratically divergent. If the SM ceases to
be applicable at a scale $\Lambda$, the mass of the Higgs boson would,
therefore, be driven to the same order $\Lambda$. That this cannot be
so is known from the fact that this would result in a strongly-interacting
scalar sector where perturbation theory would break down. One is,
therefore, driven to argue that the tree-level mass of the Higgs
boson must cancel with the radiatively-induced self-energy function to
yield acceptable values of the physical mass of the Higgs boson
(60 GeV -- 1 TeV). Taking $\Lambda$ to be the symmetry-breaking
scale of Grand Unified Theories (GUTs), {\it i.e.} $\Lambda \sim
10^{16}$ GeV, this implies an unnatural cancellation of about
26 -- 28 orders of magnitude.

The fine-tuning problem described above
affects the masses of scalars only, since the masses of fermions
and vector bosons are protected by chiral and gauge symmetries,
so that their radiative corrections can have only logarithmic divergences.
This can be clearly seen on computation of radiative
corrections to, say, the mass of the $Z$-boson, where the quadratic
divergences in individual diagrams will cancel in the final result.
With this idea in mind, an elegant restatement of the fine-tuning
solution to the problem of runaway corrections to scalar masses
is the so-called {\it Veltman condition} \cite{veltman}. Assuming that the
underlying theory has some yet-to-be-discovered symmetry which
protects the scalar mass, one simply sets to zero the sum of computed
quadratic divergences in the radiative corrections to the scalar
self-energy. Clearly this implies some relation between the
physical Higgs boson mass and the masses of other particles such
as the top quark and the gauge bosons. The {\it explanation} of such
a relationship must lie, as already stated, in the underlying theory.
For a phenomenological study, however, the Veltman condition is
very useful, since it reduces, to some degree, the arbitrariness
in the choice of top quark and Higgs boson masses.

Application of the Veltman condition to a model implies a little
more than mere cancellation of the coefficients of the quadratic
divergences in self-energy diagrams of the scalars. The condition
must not change with renormalisation group (RG) flow of the couplings.
Thus, if $f(g_i,m_i)$ be the net coefficient of the quadratic divergence
in question, then the Veltman condition is
\begin{equation}
f(g_i,m_i) \sim \frac{v^2}{\Lambda^2}     \label{VC}
\end{equation}
where $v = \langle 0 \mid H^0 \mid 0 \rangle$. Stability
under RG flow requires
\begin{equation}
\frac{d}{dt} f(g_i,m_i) = 0  \label{VCRG}
\end{equation}
where $t \equiv \ln (\frac{Q^2}{\mu^2})$.
These two equations would lead to a unique prediction for
$m_t$, $m_H$ in the SM. Unfortunately one does not obtain any real
solution to these equations.

One solution to the fine-tuning problem lies in banishing fundamental
scalars from the theory altogether. Attempts have been made in this
direction, but without conspicuous success. The other solution to
this dilemma seems to lie in extension of the SM beyond its minimal
particle content. Typical of such solutions is
supersymmetry, where pairing of bosons and fermions occurs in such a
way that contributions to $f(g_i,m_i)$ cancel pairwise for every
SM particle and its superpartner. Even the minimal supersymmetric
extension of the SM, however, requires the addition of 66 particles
to the SM!  It is desirable, therefore, to consider minimal extensions
of the SM particle spectrum to see if the Veltman condition can be
satisfied more economically.

In this paper, we first consider the addition of vector singlet and doublet
fermions to the SM and show that this extension also fails to yield
a real solution.  However, the
further addition of a singlet real scalar (which has no interactions
with the gauge sector but interacts with the Higgs doublet and the
vectorlike fermions) not only can satisfy the Veltman condition (for
both doublet and singlet), but also leads to potentially interesting
predictions from a phenomenological point of view.

In Section 2, we discuss the Veltman condition in the SM and show
that it has  no real solution.  We also
show that inclusion of vector singlet or doublet fermions does not
improve the situation. Section 3 is devoted to a model with a
singlet real scalar and vectorlike exotic fermions. Finally, our
conclusions are given in Section 4. \\
\bigskip

\begin{center}
{\Large\bf 2. Veltman Condition in the SM}
\end{center}
\bigskip

In this work, we consider the coefficients of quadratic divergences
generated at the one-loop level, anticipating that contributions from
higher orders will be suppressed by powers of the coupling constants.
To this order, then, the Veltman condition has the most general
form
\begin{equation}
\mid m_H^2 + 2 m_W^2 + m_Z^2 - 4 m_t^2 \mid \leq \frac{16 \pi^2}{3\Lambda^2}
v^2 m_H^2    \label{VCSM0}
\end{equation}
In the limit $\Lambda \gg v$, this leads to a simple relation
\begin{equation}
m_H^2 \simeq 4 m_t^2 - 2m_W^2 - m_Z^2 \label{VCSM}
\end{equation}
which yields $m_H = 182 \pm 22$ GeV for $m_t = 174 \pm 17$ GeV.
In determining the above, we set $m_W = 80.2$ GeV, $m_Z = 91.2$ GeV.
If we allow new physics to appear at a lower scale, say 10 TeV,
in which case the right side of equation (\ref{VCSM0}) is of the
order of $m_H^2$, the uncertainty in $m_H$ is increased by about 5
GeV either way.

We can rewrite equation (\ref{VCSM}) using the tree-level relations
between masses and coupling constants in the SM. This leads to the
alternative form
\begin{equation}
8 \lambda + g_1^2 + 3 g_2^2 - 8 g_t^2 \simeq 0 \label{VCSMCC}
\end{equation}
where $m_H^2 = 2 \lambda v^2$, $m_t = g_t v/\sqrt{2}$. If we
now impose RG stability on this equation, we demand
\begin{equation}
\frac{d}{dt} [ 8 \lambda + g_1^2 + 3 g_2^2 - 8 g_t^2 ] = 0. \label{VCSMRG}
\end{equation}
Using the well-known $\beta$-functions of the SM, {\it viz.}
\begin{eqnarray}
16 \pi^2 \frac{d\lambda}{dt} & = & 12 \lambda^2+6g_t^2\lambda-{3\over 2}
g_1^2\lambda-{9\over 2}g_2^2\lambda-3g_t^4+{3\over 16}g_1^4+{3\over 8}
g_1^2g_2^2+{9\over 16}g_2^4,  \\
16 \pi^2 \frac{dg_t}{dt} & = & \Big({9\over 4}g_t^2-{17\over 24}g_1^2-
{9\over 8}g_2^2-4g_3^2\Big)g_t,  \\
16 \pi^2 \frac{dg_1}{dt} & = & {41\over 12}g_1^3,  \\
16 \pi^2 \frac{dg_2}{dt} & = & -{19\over 12}g_2^3,  \\
16 \pi^2 \frac{dg_3}{dt} & = & -{7\over 2}g_3^3 \theta (q^2-m_t^2)
-{23\over 6}g_3^3 \theta (m_t^2-q^2)
\end{eqnarray}
we obtain
\begin{eqnarray}
&{ }& 72\lambda^2+36g_t^2\lambda-45g_t^4
-9g_1^2\lambda-27g_2^2\lambda \nonumber \\
&{ }& +{25\over 4}g_1^4-{15\over 4}g_2^4+{9\over 4}g_1^2g_2^2+48g_3^2g_t^2
+{17\over 2}g_1^2g_t^2+{27\over 2}g_2^2g_t^2
=0,
                         \label{VCSMRG1}
\end{eqnarray}
Numerical studies show that equations (5,12) have no
real solutions for $m_t$, $m_H$ in the range $10 ~{\rm GeV} < m_t
< 2$ TeV. This tells us that even if the Veltman condition
is satisfied at a low energy scale, it is not valid when we go to
high energies, where the problem of runaway scalar masses reappears.

Some authors \cite{ma} have argued that $g_3$ should not appear in
the above analysis, since mass generation is essentially an
electroweak phenomenon. We do not agree with this point of view,
as $g_3$ appears only in the RG evolution of $g_t$, where its
role is  known to be important. In any case, exclusion of
$g_3$ does not improve matters significantly
\footnote{One gets a real solution $m_t=117$ GeV, a
value more or less ruled out by the CDF data \cite{abe}.}.
We also note, in passing, that
even if one considers a lower value of $\Lambda$ one does not
obtain real solutions, though, in this case, the fine-tuning
problem is not so severe.

Let us now consider an extension of the SM particle spectrum
by a single generation of exotic vectorlike singlet or doublet
fermions. We have not discussed an extra sequential generation,
or a generation of mirror fermions, since these are
severely constrained by electroweak precision tests at LEP \cite{gautam}.
Vectorlike singlets are not at all constrained by these data,
while doublets are merely constrained by the oblique parameter
$T$ to be nearly mass-degenerate. A lower bound on the masses
of vectorlike fermions from LEP data is 45 GeV, while an analysis
of CDF data tells us that  vectorlike quarks  must be heavier
than 90 GeV \cite{mukhopadhyaya}. However, these masses play no role
in the subsequent discussion.

To study the Veltman condition taking these fermions into account,
one notes that they can have gauge-invariant mass terms and can
also couple to the SM gauge bosons according to their quantum
number assignments. Taking these into account, one now obtains
$\beta$-functions
\begin{eqnarray}
16 \pi^2 \frac{dg_1}{dt} & = & {187\over 36}g_1^3,  \\
16 \pi^2 \frac{dg_2}{dt} & = & -{19\over 12}g_2^3,  \\
16 \pi^2 \frac{dg_3}{dt} & = & -{17\over 6}g_3^3 \theta (q^2-m_t^2)
-{19\over 6}g_3^3\theta (m_t^2-q^2)
\end{eqnarray}
for vector singlets and
\begin{eqnarray}
16 \pi^2 \frac{dg_1}{dt} & = & {139\over 36}g_1^3,  \\
16 \pi^2 \frac{dg_2}{dt} & = & -{1\over 4}g_2^3,  \\
16 \pi^2 \frac{dg_3}{dt} & = & -{17\over 6}g_3^3 \theta (q^2-m_t^2)
-{19\over 6}g_3^3\theta (m_t^2-q^2)
\end{eqnarray}
for vector doublets. Consequently, equation (\ref{VCSMRG1}) gets
modified to
\begin{eqnarray}
&{ }& 72\lambda^2+36g_t^2\lambda-45g_t^4
-9g_1^2\lambda-27g_2^2\lambda \nonumber \\
&{ }& +{107\over 12}g_1^4-{15\over 4}g_2^4+{9\over 4}g_1^2g_2^2+48g_3^2g_t^2
+{17\over 2}g_1^2g_t^2+{27\over 2}g_2^2g_t^2
=0,
                             \label{VCVSRG}
\end{eqnarray}
and
\begin{eqnarray}
&{ }& 72\lambda^2+36g_t^2\lambda-45g_t^4
-9g_1^2\lambda-27g_2^2\lambda \nonumber \\
&{ }& +{83\over 12}g_1^4+{9\over 4}g_2^4+{9\over 4}g_1^2g_2^2+48g_3^2g_t^2
+{17\over 2}g_1^2g_t^2+{27\over 2}g_2^2g_t^2
=0,
                             \label{VCVDRG}
\end{eqnarray}
respectively.

Numerical studies of the above equation again reveal that there
is no real solution for $m_t$, $m_H$ as in the SM. This is true
even if we consider more than one extra generation since the system of
equations changes little unless the number of extra generations
is very large. We conclude, therefore, that mere inclusion of
vectorlike fermions does not provide a solution to the
fine-tuning problem.

\bigskip

\begin{center}
{\Large\bf 3. The Singlet Higgs Boson Option}
\end{center}
\bigskip

Let us now consider the minimal extension of the SM scalar sector
by a singlet real scalar, $h^0$, which has all
$SU(3)_c \times SU(2)_L \times U(1)_Y$ quantum numbers equal
to zero and hence does not couple with any of the gauge bosons of the
SM. Thus the presence of $h^0$ does not change eqs. (9-11).

We have made three assumptions about the scalar potential. First, the
potential is bounded from below, which is, strictly speaking, a
necessary requirement and not an assumption. Second, $h^0$ and $H^0$,
the SM Higgs boson, do not mix with each other
\footnote{If they do, some
quantitative results may change but no qualitative change of what we
will discuss takes place.}. Third, $h^0$ does not have a vacuum expectation
value (VEV). As a VEV of $h^0$ will not affect the masses of the
sequential fermions and the gauge bosons, by keeping it equal to zero
we are not losing any generality. The last condition allows us to write
a term of the form $\tilde m^2 h^2$ in the scalar potential. Thus, we
can write the full potential for doublet $\Phi$ and singlet $h$ as
\begin{equation}
{\cal V}_{scalar}=-m^2\Phi^{\dag}\Phi+\lambda(\Phi^{\dag}\Phi)^2+
\tilde m^2 h^2+\tilde\lambda h^4+a (\Phi^{\dag}\Phi)h^2 \label{POT}
\end{equation}
which immediately gives
\begin{equation}
m_h^2=2\tm +av^2.
\end{equation}
The RG equations for $\lambda$, $\tl$ and $a$ are
\begin{eqnarray}
16 \pi^2 \frac{d\lambda}{dt} & = & 12 \lambda^2+6g_t^2\lambda-{3\over 2}
g_1^2\lambda-{9\over 2}g_2^2\lambda-3g_t^4+6a^2+{3\over 16}g_1^4+{3\over 8}
g_1^2g_2^2+{9\over 16}g_2^4,  \\
16\pi^2\frac{d\tl}{dt}& = & 36\tl ^2+a^2, \label{LTRG}\\
16\pi^2{da\over dt}& = & \Big( 36\lambda+72\tl+6g_t^2-{3\over 2}g_1^2-
{9\over 2}g_2^2\Big)a. \label{ARG}
\end{eqnarray}
Using this equation in the Veltman condition for the singlet field
\begin{equation}
3\tl +a =0 \label{SHVC}
\end{equation}
we get the RG stability condition
\begin{equation}
a \Big[a - (4\lambda+{2\over 3}g_t^2-{1\over 6}g_1^2-{1\over 2}g_2^2)
\Big] = 0 \label{SHRG}
\end{equation}
Eqs. (\ref{SHVC}) and (\ref{SHRG}) have no nontrivial solution for $m_t
> 102$ GeV, when the quantity in parentheses becomes positive.
When coupled also with the Veltman condition and the RG
stability condition of the SM Higgs, the set of equations have no real
solutions. Thus, the inclusion of just a singlet scalar field cannot
make the fine-tuning problem vanish at all scales upto $\Lambda$.

Let us now introduce the singlet real scalar field in conjunction with
vectorlike exotic fermions ($F$). A phenomenological study \cite{ARCSRC}
of such models reveals that there is no real bound from LEP-1 data
on the masses and couplings of the scalar and exotic fermions.
The Yukawa Lagrangian is modified to
\begin{equation}
{\cal L}_{Y}^{exotic}=-\zeta_Fh\bar F F. \label{EXYUK}
\end{equation}
This introduces four new parameters in our analysis, {\em viz.} $\zeta_N$,
$\zeta_E$, $\zeta_U$ and $\zeta_D$ where the suffixes are self-explanatory.
For simplicity, we will take $\zeta_N=\zeta_E=\zeta_U=\zeta_D
=\zeta$. We have checked that the predictions do not change if we relax
this assumption.

As the vector fermions do not couple with the SM Higgs boson, the Veltman
condition for $H^0$ will read
\begin{equation}
6\lambda+a+{3\over 4}g_1^2+{9\over 4}g_2^2-6g_t^2=0,
\end{equation}
while eq. (26) will be modified to
\begin{equation}
3\tl +a -b\zeta^2=0
\end{equation}
where $b=8$ under the assumption that all $\zeta$'s are equal. Equations
(\ref{LTRG}) and (\ref{ARG}) will be modified to
\begin{eqnarray}
16\pi^2{d\tl\over dt}& = & 36\tl ^2+a^2+4b\tl \zeta^2-b\zeta^4,\\
16\pi^2{da\over dt}& = & \big( 36\lambda+72\tl +6g_t^2+4b\zeta^2-{3\over
2}g_1^2-{9\over 2}g_2^2\Big)a.
\end{eqnarray}
The RG stability equations are
\begin{eqnarray}
&{ }& 72\lambda^2+36g_t^2\lambda-45g_t^4+36a^2+36a\lambda+72a\tl +6g_t^2a
-9g_1^2\lambda-27g_2^2\lambda+\alpha_1g_1^4 \nonumber \\
&{ }& +\alpha_2g_2^4+{9\over 4}g_1^2g_2^2+48g_3^2g_t^2
+{17\over 2}g_1^2g_t^2+{27\over 2}g_2^2g_t^2-{3\over 2}ag_1^2-{9\over 2}
ag_2^2+4b\zeta^2 a=0,\\
&{ }& 12(36\tl ^2+a^2+4b\tl\zeta^2)+4(36a\lambda+72a\tl +6g_t^2
a+4b\zeta^2a)-30b\zeta^4 -{3\over 2}ag_1^2-{9\over 2}ag_2^2=0,\nonumber\\
\end{eqnarray}
where $\alpha_1=107/12 (83/12)$ and $\alpha_2=-15/4 (9/4)$ for vector
singlet (doublet) exotic fermions.

Our results are shown in Table 1 for singlet fermions and Table 2
for doublet fermions. One notes that here we can simultaneously
solve four equations at some particular point of the $\lambda, \tl$ space
for a given $m_t$. $m_H$ lies in the range currently favoured by the
electroweak precision tests.  $a$ comes out to be negative but the potential
still remains bounded from below. It is also noteworthy that the solution
comes out in the perturbative domain of the couplings, {\em i.e.},
$|\lambda|, |\tl |, |a|, |\zeta^2|\leq 4\pi$.

An easy way to see whether any change occurs in the
result if we relax the assumption
on the equality of all the $\zeta$'s, we take (i) $\zeta_N=\zeta_E=2\zeta_U
=2\zeta_D$ and (ii) $\zeta_U=\zeta_D=2\zeta_N=2\zeta_U$. For case (i),
we have to put $b=10$ in all the abovementioned equations, and for  case
(ii), the required value is $b=14$. The solutions are observed to be
unchanged. One can carry out other checks in a similar way, with the same
result.

\bigskip
\begin{center}
{\Large\bf 4. Conclusions}
\end{center}

We have shown that the Veltman condition together
with its RG stability fails to produce any acceptable
solution in the SM. This leads us to extend the SM, first in the fermionic
sector by introducing vectorlike exotic fermions, and then in the scalar
sector by introducing a singlet real scalar. However, both of these
extensions individually fail to
produce any real solution to the Veltman condition. When we
consider both exotic fermions as well as a singlet scalar in the
particle spectrum, not only do we get solutions to the Veltman conditions
for the two scalars, but we also get a prediction of $m_H$, which is  in
the experimentally favoured range. The couplings also come out to be
perturbative in nature, which is essential for the self-consistency
of the entire scheme. This appears to be an encouraging
result which should motivate searches for singlet Higgs bosons and
exotic vectorlike fermions at the upcoming colliders.

\vskip 0.5 true cm
\begin{center}
{\large \bf {Acknowledgements}}
\end{center}

This work was partially supported by the Department of Science and
Technology, India (DO No. SR/SY/P-08/92). AK acknowledges the
hospitality of the Theoretical Physics Group of
the Tata Institute of Fundamental Research, Bombay.

\newpage
\begin{center} {\large\bf Table Captions} \end{center}

{\bf Table 1}\\ The predicted parameters for vector singlet
fermions with $m_t$ as input. All masses are in GeV.\\
\vskip 20pt
{\bf Table 2}\\ The predicted parameters for vector doublet
fermions with $m_t$ as input. All masses are in GeV.\\

\newpage
\begin{center}
{\bf Table 1} \\[5mm]
\begin{tabular}{|c|ccccc|}
\hline
$m_t$& $\lambda$& $\tl$& $a$& $\zeta$& $m_H$\\
\hline
& & & & & \\
150& 0.78& 1.06& -1.26& 0.49& 307\\
155& 0.84& 1.11& -1.33& 0.50& 319\\
160& 0.90& 1.16& -1.37& 0.51& 330\\
165& 0.96& 1.22& -1.41& 0.53& 341\\
170& 1.03& 1.27& -1.50& 0.54& 353\\
175& 1.10& 1.33& -1.58& 0.55& 365\\
180& 1.16& 1.39& -1.59& 0.57& 375\\
185& 1.24& 1.45& -1.70& 0.58& 387\\
190& 1.31& 1.51& -1.75& 0.59& 398\\
195& 1.38& 1.58& -1.79& 0.61& 409\\
200& 1.46& 1.64& -1.88& 0.62& 420\\
& & & & &  \\
\hline
\end{tabular}
\newpage

{\bf Table 2} \\[5mm]
\begin{tabular}{|c|ccccc|}
\hline
$m_t$& $\lambda$& $\tl$& $a$& $\zeta$& $m_H$\\
\hline
& & & & & \\
150& 0.78& 1.07& -1.27& 0.49& 307\\
155& 0.84& 1.12& -1.33& 0.50& 319\\
160& 0.90& 1.17& -1.37& 0.52& 330\\
165& 0.96& 1.23& -1.41& 0.53& 341\\
170& 1.03& 1.28& -1.50& 0.54& 353\\
175& 1.10& 1.34& -1.58& 0.55& 365\\
180& 1.16& 1.40& -1.59& 0.57& 375\\
185& 1.24& 1.46& -1.70& 0.58& 387\\
190& 1.31& 1.52& -1.75& 0.59& 398\\
195& 1.38& 1.59& -1.79& 0.61& 409\\
200& 1.46& 1.65& -1.88& 0.62& 420\\
& & & & & \\
\hline
\end{tabular}
\end{center}

\end{document}